\documentclass[runningheads,citeauthoryear]{apinv}
\usepackage{epsfig,cite,graphics}
\usepackage{marvosym} 
\usepackage[utf8]{inputenc}
\usepackage{cite}
\usepackage{textcomp}
\usepackage{gensymb}
\usepackage{booktabs}
\usepackage[dvipsnames]{xcolor}
\usepackage{graphicx}
\usepackage{xcolor}
\usepackage{subfigure}
\usepackage[hidelinks,breaklinks]{hyperref} 
\usepackage{multirow}
\usepackage{amsmath}
\usepackage{breqn}
\usepackage{amssymb}
\usepackage{url}
\usepackage{breakurl}

\begin{document}

\title{Fundamentals of Differential and All-Sky Aperture\\
Photometry Analysis for an Open Cluster}
\titlerunning{ Fundamentals of Aperture Photometry}
\author{Kanwar Preet Kaur\inst{}, Pankaj S. Joshi\inst{}}
\authorrunning{KP Kaur, PS Joshi}
\tocauthor{Kanwar Preet Kaur, Pankaj S. Joshi} 
\institute{International Center for Cosmology, Charotar University of Science and Technology, Gujarat 388421, India
	\newline
	\email{kanwarpreet27@gmail.com}    }
\papertype{Submitted on xx.xx.xxxx; Accepted on xx.xx.xxxx}	
\maketitle

\begin{abstract}         
This article provides detailed description on the fundamentals of aperture photometry analysis. The differential and all-sky aperture photometry techniques are described thoroughly to depict the difference between the two techniques and their selection for determining the stars' magnitudes and their respective magnitude errors. The crucial calibration parameters required for the all-sky photometry analysis such as atmospheric extinction-coefficient, air-mass, zero point, color term and color index are discussed comprehensively with their extraction from the Sloan Digital Sky Survey (SDSS) archive. The all-sky aperture photometry technique is applied on the stars of an open cluster NGC 2420 to determine their calibrated magnitudes and magnitude errors in the \textit{g}, \textit{r}, and \textit{i} bands. The images required for the analysis are extracted from data release DR12 of SDSS III archive. Herein, the photometry analysis is performed by the Makali'i: SUBARU Image Processor, a Windows-based software. This software has a simple yet effective GUI and it provides the starlight minus the background sky light value with a single click. This article would aid in providing the insight into the physics of aperture photometry by manually scanning the astronomical images. In addition, the \textit{g}, \textit{r}, and \textit{i} magnitudes are transformed to \textit{B}, \textit{V}, and \textit{R} band magnitudes of Johnson-Cousins UBVRI photometric system. The color magnitude diagram for both the standard photometry systems are also provided.
\end{abstract}

\section*{Introduction}
Astronomical images are significant in the field of astronomy as various parameters of the celestial objects could be determined from them by performing different analysis at multi- and/or broad-wavelengths, say from radio wave to gamma rays. These astronomical images are investigated through techniques such as digital image processing and analysis, photometry, astrometry, spectrometry, and polarimetry \cite{r1}, which would help in determining various phenomena such as shapes and formation of star, distance, temperature, life time (age) of a star, metallicity, supernova explosion, structure of galaxies, gamma ray bursts, activity in galactic nucleus, presence of black holes/naked singularity, presence of dark matter/dark energy, and what stars are made up of, among others. The astronomical images have digital file format known as flexible image transport system (FITS) \cite{r2}. Astronomical data taken by the scientists at different wavelengths are stored, transmitted and processed in FITS digital file format which includes multi-dimensional arrays, images, table, and quantitative information to name a few. The latest version of the FITS format, standard and other related information could be obtained from the “FITS Support Office” at https://fits.gsfc.nasa.gov/. 

The most significant information of any celestial object is its energy that is received at the receiving station as an electromagnetic radiation. The observational science of studying the amount of energy received from any celestial object is known as photometry. The astronomical photometry \cite{r3} is one which involves the study of an astronomical object by measuring the energy flux or intensity of the light coming from it. The brightness of any celestial object is measured with respect to its ‘Magnitude’ \cite{r3, r4} and if the magnitude of celestial body is known then most of the stellar parameters could be determined. There are different astronomical photometry techniques, viz., aperture photometry, point-spread-function (PSF) photometry and image subtraction, through which photometry analysis is performed \cite{r4,r5, r6}. PSF photometry is utilized for the analysis of crowded star fields- the globular galactic clusters \cite{r7}. On the other hand, the aperture photometry is considered for analyzing the open galactic clusters \cite{r7}. Further, these two photometry techniques utilizes either differential or all-sky photometry to determine the magnitude of the celestial object \cite{r1}.

Herein, an old open cluster NGC 2420 \cite{r21, r8, r9} is selected to provide the details on the procedure of performing differential and all-sky aperture photometry for the amateur astronomers. The open cluster is selected because it is crucial for investigating the stellar evolution, age, metallicity, temperature, and mass of the Galactic disc \cite{r11}. Many articles are presented in the literature which provides the photometry analysis of the open clusters. These analysis are based on "IRAF Data Reduction and Analysis System" \cite{r13, r14} which is a Linux based analysis procedure. Another approach to perform photometry is to use Python based package “ASTROPY” \cite{r15}. For the non-professionals these prevailing techniques could be difficult to understand and perform as well as to install them. Hence, this article presents the photometry analysis by utilizing a Windows-based software, viz., Makali'i: SUBARU Image Processor \cite{r12}. This software has simple yet effective GUI to perform aperture photometry which would help the amateurs to get the insight into the physics of differential and all-sky aperture photometry. The \textit{g}, \textit{r}, and \textit{i} magnitudes and magnitude errors of 192 stars in NGC 2420 are calculated using all-sky aperture photometry method through Makali'i software which are then compared with their respective standard catalog magnitudes and the magnitude errors. Further, the \textit{g}, \textit{r}, and \textit{i} magnitudes are transformed into the \textit{B}, \textit{V}, and \textit{R} magnitudes of the Johnson-Cousins UBVRI standard photometric system. The images utilized are extracted from the data release DR12 of the SDSS. In addition, the in-depth description on extracting the significant calibration parameters for all-sky photometry from the Sloan Digital Sky Survey (SDSS) archive is presented along with the inclusion of the criteria to select the standard stars in case of differential photometry \cite{r10}.

This paper is organized as follows: Section 2 provides the details on the  observational data. In Section 3, the aperture photometry is described in depth. Section 4 presents the work description with the Concluding remarks in Section 5.

\section*{1. Observational Data }
\begin{figure}[!htb]
  \begin{center}
    \centering{\epsfig{file=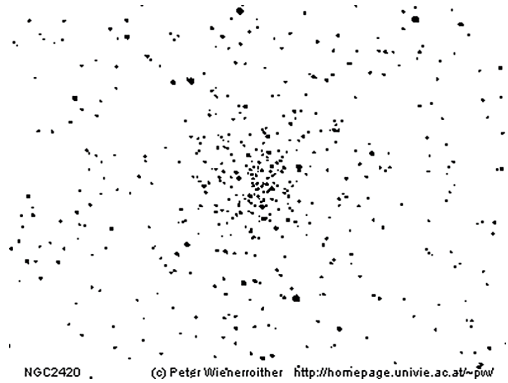, width=0.5\textwidth}}
    \caption[]{NGC 2420 (Image Credit: WEBDA)}
    \label{fig1}
  \end{center}
\end{figure}

The aperture photometry analysis presented here is performed on an open cluster NGC 2420 which is located in Gemini constellation. The other names for NGC 2420 are Cr 154 and Mel 69. The celestial coordinates of NGC 2420 taken from WEBDA (https://webda.physics.muni.cz/) are: RA (J2000) = 07h 38m 23s or 114.596\textdegree and Dec (J2000) = +21\degree\hspace{0.05cm}34\arcmin\hspace{0.05cm}24\arcsec or 21.573\textdegree. The image of NGC 2420 is presented in Fig. \,\ref{fig1} with North up and East left.

\begin{figure}[!htb]
  \begin{center}
    \centering{\epsfig{file=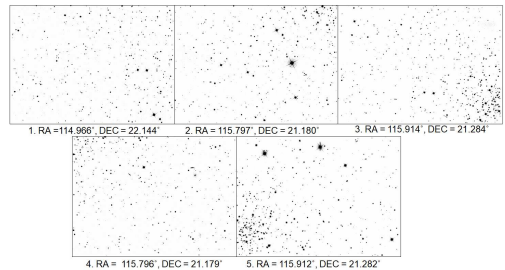, width=0.5\textwidth}}
    \caption[]{\textit{g} band images of NGC 2420 at different RA and DEC}
    \label{fig2}
  \end{center}
\end{figure}

For the all-sky photometry analysis, the calibrated images of NGC 2420 are utilized which are taken from the final Data Release 12 (DR12) of the SDSS-III (https://dr12.sdss.org/). In order to perform photometry on all the 192 stars, five images at different RA and DEC are selected under each \textit{g}, \textit{r}, and \textit{i} band. Further, the 'tsfield' file for the respective five set of images at different celestial coordinates are taken to extract the parameters for the calibration of magnitudes and also for determining the corresponding magnitude errors. The SDSS archive has images taken from 2.5-m f/5 modified Ritchey-Chrétien altitude-azimuth telescope located at Apache Point Observatory, in south east New Mexico. The five images of the NGC 2420 in \textit{g} band having angular size of 10\arcmin and the exposure time of 53.907 sec are presented in  Fig. \,\ref{fig2}.

\section*{2. Aperture Photometry }
The concept of stellar magnitudes was first brought in by the Greek astronomer Hipparchus in B.C 130. He classified stars based on their visual brightness (here on wards Magnitude) on a scale that ranged from 1 to 6. According to his classification system, the brightest starts fall under first class while stars with least brightness corresponds to the sixth category. With the advent of telescope and with the realization that the human eye has a logarithmic response to light, new magnitude scale has been introduced by Norman Pogson. In this system, over an increasing magnitude scale, any particular star is 100 times brighter than its sixth higher magnitude star \cite{r16}.

The photometry involves measurement of a stars' brightness in form of magnitude from the digital image. When an image of a star is taken then that image would not only have light from the star but it also includes the light from the background sky. This causes the light from a star to spread into wider area. Thus, it is essential to distinguish starlight from the background sky light. This could be done in different ways, of which one of the most popular technique being aperture photometry. It should also to be noted that prior to measuring the magnitudes from the images various instrumental effects should be removed from them. Some of the image calibrations requirements are de-biasing, flat-fielding, and removal of the bad pixels effects and cosmic-ray events \cite{r10}.

\begin{figure}[!htb]
  \begin{center}
    \centering{\epsfig{file=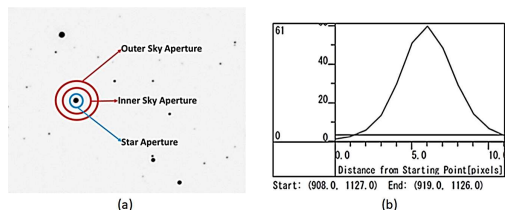, width=0.5\textwidth}}
    \caption[]{(a) Star aperture and sky annulus, and (b) Brightness profile of a star located at RA = 114.346\textdegree and DEC = 21.676\textdegree in NGC 2420 obtained through Makali'i software}
    \label{fig3}
  \end{center}
\end{figure}

In aperture photometry, pixels that contain starlight are covered by a small circular patch known as aperture. The background sky light is then captured by a doughnut shaped patch, also named as annulus, whose center radius is larger than that of the star aperture.  Fig. \,\ref{fig3}(a) shows the star aperture and the sky annulus. The light from the background sky is then eliminated by subtracting pixels in background sky annulus from the star aperture \cite{r1,r5}. The best way to select appropriate star radius is to first find the full width half maximum (FWHM) of the star brightness curve or find its Gaussian sigma {($\sigma$)}. Then the radius of star aperture is selected as $\sigma$ or half of the FWHM value. However, to capture the starlight in totality the best practice is to set star radius five times the $\sigma$ radius. Whereas, the inner radius of the background sky aperture should be set at least twice that of the star aperture. This would prevent the inclusion of any starlight into the sky annulus. The  Fig. \,\ref{fig3}(b) depicts the brightness profile of a star located at RA = 114.346\textdegree and DEC = 21.676\textdegree in NGC 2420 which is obtained through Makali'i software. The FWHM and $\sigma$ are related by the Eq. (\ref{eq_1}).

\begin{eqnarray}\label{eq_1}
    FWHM &=& 2.37\sigma
\end{eqnarray}

Once the photometric observations are achieved, they could be fitted to the theoretical isochrones to determine the essential stellar parameters such as metallicity, stellar reddening, distance and ages. The isochrones are the mathematical models which represents the Hertzsprung–Russell (HR) diagram for a particular metallicity and age \cite{r17}.  For instance, the surface temperature of an object could be determined by fitting the color-magnitude diagram (CMD) with the HR diagram. Thus, different isochrones are overlapped with the photometric observations and the best fit isochrones are selected for the determination of the above mentioned parameters. The database for isochrones could be found at the website of Padova database (http://pleiadi.pd.astro.it/).

The magnitude of a star is obtained by accessing the pixel values present in the star aperture and the sky annulus. These pixel values are also known as “pixel count” and the magnitude obtained from these pixel counts is known as “instrumental magnitude”. There are various ways to obtain instrumental magnitude from the star aperture counts and the background sky aperture count \cite{r1, r10}. One of the relationships between them is shown in Eq. (\ref{eq_2}) \cite{r1}:

\begin{eqnarray}\label{eq_2}
   m_{ins} &=& -2.5\left(\frac{C_{apt}-n_{apt}\left(\frac{C_{apt}}{n_{ann}}\right)}{t}\right)+\Lambda
\end{eqnarray}

In the relation (\ref{eq_2}), the $C_{apt}$ and $n_{apt}$ are the sum of pixel values and number of pixels in the star aperture, respectively, $n_{ann}$ is the number of pixels in the background sky annulus, and $\Lambda$ is an arbitrary constant. The arbitrary constant, $\Lambda$, is added to the instrumental magnitudes so that color difference becomes equal to zero \cite{r1, r10}.

\subsection*{2.1. Differential Aperture Photometry}
In the differential aperture photometry, the apparent magnitude of the target star is obtained from the instrumental magnitude by comparing the target star with the standard star (also known as comparison star) that is present in the same field of view (FOV). In this technique, the target star has the same FOV, same observation time, also it is observed through same filter and same atmosphere as that of the standard star. Hence, all the atmospheric and equipment parameters affecting the brightness are cancelled out when the difference between the standard and target star is taken. The only significant parameter to be taken into account is the difference of the target and the standard star. The apparent magnitude of the target star in this case is obtained by the  Eq. (\ref{eq_3}) \cite{r4}.

\begin{eqnarray}\label{eq_3}
   M_t &=& \left(m_{ins\_t}-m_c\right)+M_c
\end{eqnarray}

In the Eq. (\ref{eq_3}), $M_{t}$ and $m_{ins\_t}$ are the apparent and instrumental magnitude of the target star, respectively, while the standard and the instrumental magnitude of the standard (comparison) star is represented by $M_{c}$ and $m_{c}$, respectively. The selection of standard star is crucial while performing the differential photometry as the magnitude obtained here is solely dependent on the difference of the standard and target star. Thus, the standard star magnitude should be accurately determined in this technique. The points that should be taken care of while selecting the standard star for comparison are as follows \cite{r5, r10}:

\renewcommand{\labelenumi}{\roman{enumi}}
\begin{enumerate}
  \item It should be a non-variable star.
 \item	It should not be a saturated star
 \item	It should lie near to the target star but not near any edges of the image.
 \item	It should be a discrete star not blended with other star/s.
 \item	All the comparison stars should have similar wavelength (color).
 \item It should not be a very blue or very red star.
 \item It should have at least 100 signal-to-noise ratio (SNR) value.
 \item	 It should have the magnitude near to the target star so that the magnitude error is similar.
\end{enumerate}

Herein, the Makali'i software is utilized to obtain the star count and the background sky count pixels values for the determination of instrumental magnitude. Makali'i software provides the “Result count” by a single click on the image (\textit{g}, \textit{r}, and \textit{i} images of the NGC 2420 in FITS file format) at the location of the target/comparison star. The results generated on the software could be stored as CSV file format which is then used to obtain the magnitude of the target and the standard stars. The instrumental magnitude is obtained by substituting "Result count" in the Eq. (\ref{eq_4}). Then to this instrumental magnitude, the standard magnitude value of the comparison star (obtained from the catalog) and its pixel count are added (Ref. Eq. (\ref{eq_6})) to obtain the apparent magnitude.

\begin{eqnarray}\label{eq_4}
   m &=& -2.5\log_{10} {(count)}
\end{eqnarray}
\begin{eqnarray}\label{eq_5}
   m_{target}-m_{comp} &=& -2.5\log_{10} {\left(count_{target}\right)} -\left(-2.5log_{10} {(count_{comp})}\right)
\end{eqnarray}
\begin{eqnarray}\label{eq_6}
   m_{target} &=& -2.5log_{10}{\left(\frac{count_{target}}{count_{comp}}\right)}+m_{comp}
\end{eqnarray}

Although, the differential photometry is easy but to determine the magnitude of any target star, the standard magnitude of the comparison star should be known beforehand. Also, the comparison star should be on the same FOV as that of the target star. If the comparison star is located out of the FOV then the magnitude is obtained by the all-sky aperture photometry technique. The differential photometry is employed, in general, when it is required to investigate any change in the "target" over certain interval of time. In this technique, accuracy of the apparent magnitude depends on the selection of the comparison star and on the exactness of the target and comparison star difference.

\subsection*{2.2. All-Sky Aperture Photometry}
All-sky photometry provides the magnitude of the target directly from the images taken at known atmosphere and system from a set of standard stars. This technique is applied when the standard stars are outside the FOV, when the images are taken at different time, through different filters and atmosphere. Hence, the all-sky photometry requires meticulous observations that are subjected to thorough and careful data reductions. This section  presents the procedures, in-detail, to obtain the values for the calibration parameters, viz., atmospheric extinction-coefficient, air-mass, zero point, color term and color index. In addition, the steps to calculate the error magnitude is presented.

The procedure to find the instrumental magnitude of the star in this technique is similar to that of the differential photometry. The formula for determining the instrumental magnitude is defined by the Eq. (\ref{eq_7}).

\begin{eqnarray}\label{eq_7}
   m_{ins} &=& -2.5\log_{10} {(count_{target})}+\Lambda
\end{eqnarray}

Eventually, the instrumental magnitude is converted into calibrated magnitude. The calibrated magnitude takes into account of effects of atmospheric extinction coefficient (k), air-mass(X), color correction term (C), and color index (CI), to determine a magnitude which is tied to the standard photometric system \cite{r18}. The calibrated magnitude is defined by the Eq. (\ref{eq_8}) \cite{r1, r4, r6, r10}.

\begin{eqnarray}\label{eq_8}
   M=m_{ins}-\Lambda+ZP+kX+C\times CI
\end{eqnarray}

The following section presents the steps to obtain the values of the above mentioned calibration parameters \cite{r1, r4,r6, r10} from the observed images.

\subsubsection{2.2.1 Air-mass(X):} 
Air-mass is known as the length of the air-path  which starlight travels through. As the light travels through some distance of the atmosphere, some fixed portion of light gets attenuated. Thus, it is necessary to find out what amount of light is lossed due to a particular unit of the air-mass. The air-mass (length of air-path) depends solely on the zenith distance, $z_{L}$. The zenith distance is obtained from the Eq. (\ref{eq_9}) \cite{r10}.

\begin{eqnarray}\label{eq_9}
   X=sec(z_{L}) &=& \frac{1}{(sin\phi) (sin\delta) +(cos\phi) (cos\delta) (cosh)}
\end{eqnarray}

In the above relationship, $\phi$ is the latitude of observation, $\delta$ and $h$ are the Declination and the hour angle of the observed target, respectively. The Right Ascension ($\alpha$) of the observed target and the local sidereal time ($s$) are related to the hour angle by the Eq. (\ref{eq_10}).

\begin{eqnarray}\label{eq_10}
   h &=& s-\alpha
\end{eqnarray}

\subsubsection{2.2.2 Atmospheric Extinction Coefficient (k):} 
The atmospheric extinction is defined as the dimming of a starlight as it passes through the atmosphere. It has the units of magnitudes/air-mass. As the air-path  increases, the extinction also increases. 
 
There are various methods to attain extinction coefficient. One of the widely used method to determine the atmospheric extinction is to observe the same non-variable star during the night with any of the suitable filter at various intervals as well as at different zenith angles. The observed magnitude of the observed star is then plotted against the calculated air-mass. The graph thus, plotted should have a straight line where the slope is equal to the extinction. The atmospheric extinction depends strongly on the wavelength. For instance, extinction is lower for red band than for the blue. Thus, in order to neutralize the extinction precisely, the color of the star has to be included. The extinction is then represented by the Eq. (\ref{eq_11}).

\begin{eqnarray}\label{eq_11}
   k &=& k^{'}-k^{''} (CI)
\end{eqnarray}

$k^{'}$ is the first order extinction in the particular color filter, $k^{''}$ is the second order extinction for that filter, CI is the standard color index. $k^{''}$ has the units of magnitudes/air-mass /magnitude of CI. Theoretically, $k^{''}$ is 0 for the V filter and for the following CI: U–B and B–V. Practically, the value of $k^{''}$ is very small and is hard to measure, hence it is ignored. However, to find $k^{''}$, one could refer \cite{r4}. The magnitude after the compensation of extinction is represented with a suffix ‘o’, it indicates that extinction effects have been removed from the magnitude. Sometimes, this magnitude is also known as exoatmospheric magnitude and is represented by the Eq.(\ref{eq_12})

\begin{eqnarray}\label{eq_12}
   m_o &=& m-k^{'}X-k^{''}X(CI)
\end{eqnarray}

\subsubsection{2.2.3 Color Index (CI) and Color-Correction Term (C):} 
The system response for a particular band of color filter may not be identical to that of the standard photometric system. Thus, the color index and color correction terms are included to remove any color dependency. The color term is determined by obtaining the difference of the adjacent color filter band. For instance, B-V, V-R, g-r, and r-i, to name a few. The change in the response between the system under consideration and the standard photometric system is linear for a particular color band (Ref.  Fig. \,\ref{fig5}). Hence, the color correction term is obtained in a manner similar to that applied for obtaining $k$. That is, observing the star for certain interval of time throughout the night at different $z_L$ and then the best values for $k$ and $C$ are determined by fitting the Eq. (\ref{eq_13}) using the method of least square.

\begin{eqnarray}\label{eq_13}
   M &=& m_{ins}-\Lambda+kX+C\times{CI}
\end{eqnarray}

\subsubsection{2.2.4 Zero Point (ZP):} 
Eventually, the exoatmospheric magnitude obtained from the above calibration process is transformed to a magnitude that is tied to a standard photometric system. For the determination of the zero point it is essential to observe a primary or secondary standard star. The zero point is determined from the exoatmospheric standard star magnitude. The zero point magnitude is then calculated from Eq(\ref{eq_14}).

\begin{eqnarray}\label{eq_14}
   m_{zp} &=& m_{std}-m_{ostd}
\end{eqnarray}

In the above equation, $m_{zp}$ is the zero-point magnitude, $m_{std}$ is the catalog standard magnitude and $m_{ostd}$ is the exoatmosphric instrumental magnitude of the standard star.

\subsection*{2.3. Error Magnitude}
Because of the Poisson statistic behaviour of the detected photon there is a inherent error in the photometry magnitude. The two ways to represent this error or uncertainty are the signal-to-noise ratio and the standard deviation present in the calculated magnitude of the star. The magnitude error, $\sigma_m$, for a star is defined by the Eq(\ref{eq_15}). This formula determines that if $S/N$ is $x$, then there is $x\%$ of error in the measured magnitude \cite{r1, r25}.

\begin{eqnarray}\label{eq_15}
   \sigma_m &=& \frac{1.086}{S/N}
\end{eqnarray}

There is single signal source while there are several sources of noise such as noises from the sky as well as noises from the detector. The noise sources taken into account are readout noise, dark current, quantization noise, noise due to star background count, to name a few. The signal-to-noise ratio is thus defined by the Eq. (\ref{eq_15}) \cite{r25}.

\begin{eqnarray}\label{eq_16}
   \frac{S}{N} &=& \frac{N_*}{\sqrt{N_*+n_{pix}(N_S+N_D+N_R^2)}}
\end{eqnarray}

The numerator in the above equation, $N_{\**}$, is the total number of photons detected from the target. The terms in denominator of the Eq(\ref{eq_16}) are defined as $n_{pix}$  is the number of pixels present in the target, $N_S$ is the total number of photons per pixel from the background sky, $N_D$ is the total number of dark current electrons per pixel, and $N_R^2$ is the total number of electrons per pixel resulting from the readout noise. The further details on the magnitude error could be found in \cite{r25}.

\section*{3. WORK DESCRIPTION}
\begin{figure}[!htb]
  \begin{center}
    \centering{\epsfig{file=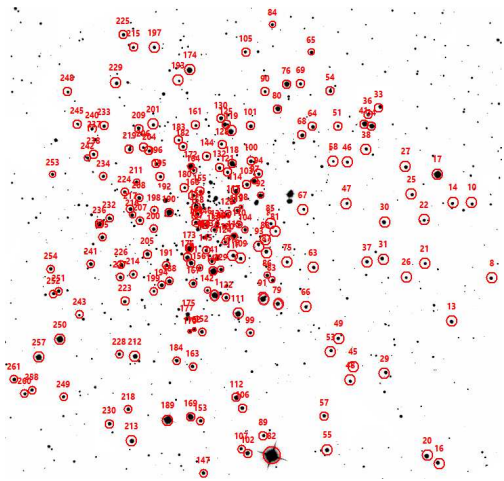, width=0.5\textwidth}}
    \caption[]{\textit{g}-band image of NGC 2420 with the 192 selected stars}
    \label{fig4}
  \end{center}
\end{figure}

The all-sky photometry is performed on the 192 stars lying within and near the radius of 6\arcmin from the center of NGC 2420. The stars are selected on the basis of brightness perceived through naked eyes. The selected region and the star numbers are adopted from \cite{r21}. As mentioned in Section 1, five different images are extracted for each \textit{g}, \textit{r}, and \textit{i} bands to cover all the 192 stars. The combined \textit{g} band image with the selected 192 stars is presented in Fig. \,\ref{fig4} (North up and East left) and their celestial coordinates are presented in Table \ref{table1}. Each of these fifteen images are then analyzed through Makali'i software to convert the raw counts of the selected stars into the raw instrumental magnitudes. As the NGC 2420 is an old open cluster, hence the values for air-mass ($X$), atmospheric extinction ($k$), and zero points ($ZP$) are obtained from the 'tsfield' file present in the SDSS archive. The values of these calibration parameters are depicted in Table \ref{table2}.

\begin{table}[htbp]
  \centering
  \caption{Celestial coordinates of the 192 selected stars in NGC 2420}
    \begin{tabular}{c c c c c c c c c c c c}
    \toprule
    \multicolumn{1}{p{2.1em}}{\centering{\textbf{Star No.}}} & 
    \multicolumn{1}{p{3.4em}}{\centering{\textbf{RA(\degree)}}} & 
    \multicolumn{1}{p{3.4em}}{\centering{\textbf{DEC(\degree)}}} & \multicolumn{1}{p{2.1em}}{\centering{\textbf{Star No.}}} & 
    \multicolumn{1}{p{3.4em}}{\centering{\textbf{RA(\degree)}}} & 
    \multicolumn{1}{p{3.4em}}{\centering{\textbf{DEC(\degree)}}} & 
    \multicolumn{1}{p{2.1em}}{\centering{\textbf{Star No.}}} & 
    \multicolumn{1}{p{3.4em}}{\centering{\textbf{RA(\degree)}}} & 
    \multicolumn{1}{p{3.4em}}{\centering{\textbf{DEC(\degree)}}} & 
    \multicolumn{1}{p{2.1em}}{\centering{\textbf{Star No.}}} & 
    \multicolumn{1}{p{3.4em}}{\centering{\textbf{RA(\degree)}}} & 
    \multicolumn{1}{p{3.4em}}{\centering{\textbf{DEC(\degree)}}} \\
    \midrule
    \textbf{8} & 114.464 & 21.544 & \textbf{84} & 114.572 & 21.661 & \textbf{138} & 114.600 & 21.567 & \textbf{199} & 114.630 & 21.539 \\
    \textbf{10} & 114.473 & 21.579 & \textbf{85} & 114.572 & 21.569 & \textbf{140} & 114.601 & 21.548 & \textbf{200} & 114.630 & 21.568 \\
    \textbf{13} & 114.483 & 21.524 & \textbf{86} & 114.574 & 21.546 & \textbf{141} & 114.601 & 21.553 & \textbf{201} & 114.630 & 21.615 \\
    \textbf{14} & 114.483 & 21.578 & \textbf{87} & 114.574 & 21.556 & \textbf{142} & 114.603 & 21.539 & \textbf{204} & 114.632 & 21.603 \\
    \textbf{16} & 114.490 & 21.459 & \textbf{88} & 114.575 & 21.561 & \textbf{143} & 114.603 & 21.564 & \textbf{205} & 114.633 & 21.556 \\
    \textbf{17} & 114.490 & 21.592 & \textbf{89} & 114.575 & 21.473 & \textbf{144} & 114.604 & 21.601 & \textbf{206} & 114.635 & 21.604 \\
    \textbf{20} & 114.495 & 21.463 & \textbf{90} & 114.575 & 21.630 & \textbf{145} & 114.604 & 21.557 & \textbf{207} & 114.637 & 21.572 \\
    \textbf{21} & 114.497 & 21.551 & \textbf{91} & 114.576 & 21.535 & \textbf{146} & 114.604 & 21.569 & \textbf{208} & 114.637 & 21.582 \\
    \textbf{22} & 114.497 & 21.571 & \textbf{92} & 114.577 & 21.583 & \textbf{147} & 114.605 & 21.455 & \textbf{209} & 114.638 & 21.613 \\
    \textbf{25} & 114.503 & 21.583 & \textbf{93} & 114.578 & 21.560 & \textbf{149} & 114.605 & 21.563 & \textbf{211} & 114.639 & 21.589 \\
    \textbf{26} & 114.506 & 21.545 & \textbf{94} & 114.578 & 21.593 & \textbf{152} & 114.605 & 21.520 & \textbf{212} & 114.639 & 21.509 \\
    \textbf{27} & 114.506 & 21.595 & \textbf{97} & 114.580 & 21.589 & \textbf{153} & 114.606 & 21.479 & \textbf{213} & 114.640 & 21.470 \\
    \textbf{28} & 114.507 & 21.480 & \textbf{99} & 114.582 & 21.519 & \textbf{155} & 114.607 & 21.585 & \textbf{214} & 114.640 & 21.547 \\
    \textbf{29} & 114.515 & 21.500 & \textbf{100} & 114.582 & 21.598 & \textbf{156} & 114.608 & 21.550 & \textbf{215} & 114.640 & 21.650 \\
    \textbf{30} & 114.517 & 21.570 & \textbf{101} & 114.582 & 21.614 & \textbf{157} & 114.608 & 21.568 & \textbf{216} & 114.641 & 21.574 \\
    \textbf{31} & 114.518 & 21.554 & \textbf{102} & 114.583 & 21.464 & \textbf{158} & 114.608 & 21.574 & \textbf{217} & 114.642 & 21.577 \\
    \textbf{33} & 114.519 & 21.623 & \textbf{103} & 114.584 & 21.588 & \textbf{159} & 114.609 & 21.578 & \textbf{218} & 114.642 & 21.484 \\
    \textbf{34} & 114.523 & 21.614 & \textbf{104} & 114.585 & 21.567 & \textbf{160} & 114.609 & 21.570 & \textbf{219} & 114.642 & 21.604 \\
    \textbf{36} & 114.524 & 21.619 & \textbf{105} & 114.585 & 21.648 & \textbf{161} & 114.609 & 21.615 & \textbf{223} & 114.644 & 21.534 \\
    \textbf{37} & 114.525 & 21.552 & \textbf{106} & 114.586 & 21.485 & \textbf{162} & 114.609 & 21.521 & \textbf{224} & 114.644 & 21.585 \\
    \textbf{38} & 114.526 & 21.603 & \textbf{107} & 114.587 & 21.467 & \textbf{163} & 114.610 & 21.504 & \textbf{225} & 114.645 & 21.656 \\
    \textbf{40} & 114.526 & 21.623 & \textbf{108} & 114.587 & 21.577 & \textbf{164} & 114.610 & 21.594 & \textbf{226} & 114.646 & 21.552 \\
    \textbf{41} & 114.526 & 21.615 & \textbf{109} & 114.587 & 21.553 & \textbf{166} & 114.611 & 21.583 & \textbf{227} & 114.646 & 21.545 \\
    \textbf{45} & 114.533 & 21.496 & \textbf{110} & 114.588 & 21.569 & \textbf{167} & 114.611 & 21.543 & \textbf{228} & 114.647 & 21.510 \\
    \textbf{46} & 114.535 & 21.598 & \textbf{111} & 114.588 & 21.528 & \textbf{169} & 114.611 & 21.480 & \textbf{229} & 114.648 & 21.634 \\
    \textbf{47} & 114.535 & 21.578 & \textbf{112} & 114.589 & 21.490 & \textbf{170} & 114.611 & 21.520 & \textbf{230} & 114.651 & 21.478 \\
    \textbf{48} & 114.535 & 21.493 & \textbf{113} & 114.589 & 21.576 & \textbf{171} & 114.612 & 21.552 & \textbf{232} & 114.652 & 21.573 \\
    \textbf{49} & 114.539 & 21.516 & \textbf{114} & 114.590 & 21.585 & \textbf{172} & 114.612 & 21.596 & \textbf{233} & 114.654 & 21.615 \\
    \textbf{51} & 114.540 & 21.614 & \textbf{115} & 114.590 & 21.564 & \textbf{173} & 114.612 & 21.558 & \textbf{234} & 114.655 & 21.592 \\
    \textbf{53} & 114.542 & 21.511 & \textbf{116} & 114.590 & 21.555 & \textbf{174} & 114.613 & 21.640 & \textbf{235} & 114.655 & 21.564 \\
    \textbf{54} & 114.541 & 21.633 & \textbf{117} & 114.590 & 21.580 & \textbf{175} & 114.612 & 21.528 & \textbf{236} & 114.657 & 21.570 \\
    \textbf{55} & 114.544 & 21.466 & \textbf{118} & 114.592 & 21.598 & \textbf{176} & 114.613 & 21.554 & \textbf{237} & 114.659 & 21.608 \\
    \textbf{57} & 114.546 & 21.481 & \textbf{119} & 114.592 & 21.612 & \textbf{177} & 114.613 & 21.526 & \textbf{238} & 114.660 & 21.602 \\
    \textbf{58} & 114.547 & 21.596 & \textbf{120} & 114.593 & 21.563 & \textbf{180} & 114.615 & 21.586 & \textbf{240} & 114.660 & 21.613 \\
    \textbf{63} & 114.551 & 21.549 & \textbf{121} & 114.594 & 21.593 & \textbf{182} & 114.616 & 21.605 & \textbf{241} & 114.661 & 21.551 \\
    \textbf{64} & 114.552 & 21.614 & \textbf{122} & 114.593 & 21.535 & \textbf{183} & 114.617 & 21.608 & \textbf{242} & 114.663 & 21.600 \\
    \textbf{65} & 114.552 & 21.648 & \textbf{123} & 114.594 & 21.574 & \textbf{184} & 114.618 & 21.507 & \textbf{245} & 114.667 & 21.615 \\
    \textbf{66} & 114.555 & 21.531 & \textbf{124} & 114.595 & 21.561 & \textbf{188} & 114.622 & 21.544 & \textbf{248} & 114.672 & 21.630 \\
    \textbf{67} & 114.556 & 21.576 & \textbf{125} & 114.595 & 21.615 & \textbf{189} & 114.622 & 21.479 & \textbf{249} & 114.674 & 21.491 \\
    \textbf{68} & 114.557 & 21.610 & \textbf{126} & 114.595 & 21.568 & \textbf{190} & 114.622 & 21.575 & \textbf{250} & 114.676 & 21.517 \\
    \textbf{69} & 114.558 & 21.634 & \textbf{127} & 114.596 & 21.538 & \textbf{191} & 114.624 & 21.551 & \textbf{251} & 114.676 & 21.539 \\
    \textbf{75} & 114.564 & 21.552 & \textbf{128} & 114.597 & 21.606 & \textbf{192} & 114.625 & 21.581 & \textbf{252} & 114.679 & 21.538 \\
    \textbf{76} & 114.564 & 21.634 & \textbf{129} & 114.597 & 21.549 & \textbf{193} & 114.618 & 21.635 & \textbf{253} & 114.680 & 21.593 \\
    \textbf{79} & 114.568 & 21.533 & \textbf{130} & 114.597 & 21.618 & \textbf{194} & 114.626 & 21.542 & \textbf{254} & 114.680 & 21.549 \\
    \textbf{80} & 114.569 & 21.623 & \textbf{132} & 114.598 & 21.595 & \textbf{195} & 114.627 & 21.591 & \textbf{257} & 114.686 & 21.509 \\
    \textbf{81} & 114.570 & 21.566 & \textbf{133} & 114.598 & 21.570 & \textbf{196} & 114.629 & 21.597 & \textbf{258} & 114.689 & 21.494 \\
    \textbf{82} & 114.571 & 21.463 & \textbf{134} & 114.599 & 21.568 & \textbf{197} & 114.650 & 21.650 & \textbf{260} & 114.693 & 21.492 \\
    \textbf{83} & 114.571 & 21.543 & \textbf{135} & 114.599 & 21.554 & \textbf{198} & 114.630 & 21.576 & \textbf{261} & 114.698 & 21.499 \\
    \bottomrule
    \end{tabular}%
  \label{table1}%
\end{table}%

\begin{table}[htbp]
  \centering
  \caption{Calibration parameters obtained from the 'tsfield' files of the respective five different set of images}
    \begin{tabular}{cccc}
    \toprule
    \multicolumn{1}{p{4em}}{\centering{\textbf{Set of Images}}} & 
    \multicolumn{1}{p{5em}}{\centering{\textbf{air mass}}} & \multicolumn{1}{p{6em}}{\centering{\textbf{Zero Point}}} & \multicolumn{1}{p{9em}}{\centering{\textbf{Extinction Factor}}} \\
    \midrule
    \textbf{1} & 1.2068 & 23.818 & 0.414 \\
    \textbf{2} & 1.1815 & 23.507 & 0.4176 \\
    \textbf{3} & 1.1805 & 23.507 & 0.4176 \\
    \textbf{4} & 1.0657 & 23.521 & 0.4467 \\
    \textbf{5} & 1.0651 & 23.526 & 0.4467 \\
    \bottomrule
    \end{tabular}%
  \label{table2}%
\end{table}%

The final calibrated magnitudes and the magnitude errors for the selected stars in \textit{g}, \textit{r}, and \textit{i} bands are obtained by substituting the values of the calibration parameters from the Table \ref{table2} into the relations described in Section 2.2 and Section 2.3, respectively. The excel spreadsheets are utilized for finding out the results from the respective equations. The SAOImage DS9 \cite{r20} software is used to locate and assign numbers to the selected stars. The calculated calibrated magnitudes and error magnitudes for the first 30 stars are depicted in Table \ref{table3}, which are also compared with their respective catalog values. The SDSS DR12 catalog \cite{r24} is referred for comparing the calculated results.

\begin{table}[htbp]
  \centering
  \caption{Calculated calibrated magnitudes and error magnitudes for the first 30 stars in \textit{g}, \textit{r}, and \textit{i} bands}
    \begin{tabular}{ccccccccccccc}
    \toprule
    \multirow {2}{2em}{\centering{\textbf{Stars}}} &      \multicolumn {3}{p{6em}} {\centering{\textbf{Calculated Calibrated Magnitudes}}} &        
    \multicolumn {3}{p{6em}} {\centering{\textbf{Calculated Magnitude Error}}} &                 
    \multicolumn {3}{p{6em}} {\centering{\textbf{Catalog Magnitudes}}} &        
    \multicolumn {3}{p{6em}} {\centering{\textbf{Catalog Magnitudes Errors}}}  \\
    \cmidrule{2-13}   & 
    \multicolumn{1}{p{2em}}{\centering{\textit{\textbf{g}}}} & \multicolumn{1}{p{2em}}{\centering{\textit{\textbf{r}}}} & \multicolumn{1}{p{2em}}{\centering{\textit{\textbf{i}}}} & \multicolumn{1}{p{2em}}{\centering{\textit{\textbf{e\_g}}}} & \multicolumn{1}{p{2em}}{\centering{\textit{\textbf{e\_r}}}} & \multicolumn{1}{p{2em}}{\centering{\textit{\textbf{e\_i}}}} & \multicolumn{1}{p{2em}}{\centering{\textit{\textbf{g}}}} & \multicolumn{1}{p{2em}}{\centering{\textit{\textbf{r}}}} & \multicolumn{1}{p{2em}}{\centering{\textit{\textbf{i}}}} & \multicolumn{1}{p{2em}}{\centering{\textit{\textbf{e\_g}}}} & \multicolumn{1}{p{2em}}{\centering{\textit{\textbf{e\_r}}}} & \multicolumn{1}{p{2em}}{\centering{\textit{\textbf{e\_i}}}} \\
    \midrule
    \textbf{8} & 13.695 & 13.823 & 14.070 & 0.0014 & 0.0015 & 0.0016 & 14.866 & 14.653 & 14.608 & 0.003 & 0.004 & 0.004 \\
    \textbf{10} & 14.408 & 14.422 & 14.617 & 0.0019 & 0.0019 & 0.0021 & 15.566 & 15.245 & 15.143 & 0.003 & 0.004 & 0.004 \\
    \textbf{13} & 13.437 & 13.261 & 14.128 & 0.0012 & 0.0011 & 0.0017 & 14.663 & 14.206 & 14.241 & 0.003 & 0.008 & 0.005 \\
    \textbf{14} & 14.165 & 14.066 & 14.217 & 0.0017 & 0.0016 & 0.0017 & 15.332 & 14.902 & 14.747 & 0.003 & 0.004 & 0.004 \\
    \textbf{16} & 13.147 & 13.277 & 13.943 & 0.0011 & 0.0011 & 0.0016 & 14.362 & 14.162 & 13.27 & 0.003 & 0.004 & 0.003 \\
    \textbf{17} & 12.444 & 12.842 & 13.568 & 0.0008 & 0.0009 & 0.0013 & 11.787 & 11.654 & 11.686 & 0.001 & 0.001 & 0.001 \\
    \textbf{20} & 13.355 & 13.416 & 14.282 & 0.0012 & 0.0012 & 0.0018 & 15.024 & 14.751 & 14.973 & 0.01  & 0.011 & 0.012 \\
    \textbf{21} & 14.189 & 14.202 & 14.406 & 0.0017 & 0.0017 & 0.0019 & 15.36 & 15.049 & 14.934 & 0.003 & 0.004 & 0.004 \\
    \textbf{22} & 14.361 & 13.936 & 13.963 & 0.0019 & 0.0015 & 0.0016 & 15.52 & 14.761 & 14.475 & 0.003 & 0.004 & 0.004 \\
    \textbf{25} & 13.990 & 13.819 & 13.935 & 0.0016 & 0.0014 & 0.0015 & 15.159 & 14.651 & 14.462 & 0.003 & 0.004 & 0.004 \\
    \textbf{26} & 14.472 & 14.380 & 14.533 & 0.0020 & 0.0019 & 0.0020 & 15.643 & 15.205 & 15.054 & 0.003 & 0.004 & 0.004 \\
    \textbf{27} & 13.462 & 13.552 & 13.785 & 0.0012 & 0.0013 & 0.0014 & 14.631 & 14.388 & 14.315 & 0.003 & 0.004 & 0.004 \\
    \textbf{28} & 14.333 & 14.441 & 14.685 & 0.0019 & 0.0020 & 0.0022 & 15.541 & 15.316 & 15.259 & 0.003 & 0.004 & 0.004 \\
    \textbf{29} & 13.528 & 13.256 & 14.010 & 0.0013 & 0.0011 & 0.0016 & 14.744 & 14.215 & 14.177 & 0.003 & 0.008 & 0.005 \\
    \textbf{30} & 13.105 & 13.035 & 13.983 & 0.0010 & 0.0010 & 0.0016 & 14.373 & 13.964 & 13.959 & 0.003 & 0.006 & 0.004 \\
    \textbf{31} & 13.639 & 13.675 & 13.877 & 0.0013 & 0.0014 & 0.0015 & 14.803 & 14.508 & 14.396 & 0.003 & 0.004 & 0.004 \\
    \textbf{33} & 13.466 & 13.549 & 13.787 & 0.0012 & 0.0013 & 0.0014 & 14.63 & 14.384 & 14.316 & 0.003 & 0.004 & 0.004 \\
    \textbf{34} & 12.860 & 13.217 & 13.600 & 0.0009 & 0.0011 & 0.0013 & 15.438 & 15.319 & 14.789 & 0.012 & 0.012 & 0.012 \\
    \textbf{36} & 13.833 & 13.958 & 14.213 & 0.0015 & 0.0015 & 0.0017 & 15.003 & 14.793 & 14.738 & 0.003 & 0.004 & 0.004 \\
    \textbf{37} & 13.546 & 13.226 & 13.934 & 0.0013 & 0.0011 & 0.0015 & 14.809 & 14.707 & 15.601 & 0.003 & 0.011 & 0.013 \\
    \textbf{38} & 14.063 & 14.136 & 14.358 & 0.0016 & 0.0017 & 0.0019 & 15.235 & 14.965 & 14.884 & 0.003 & 0.004 & 0.004 \\
    \textbf{40} & 14.541 & 14.621 & 14.846 & 0.0020 & 0.0021 & 0.0024 & 15.714 & 15.458 & 15.374 & 0.004 & 0.004 & 0.004 \\
    \textbf{41} & 12.813 & 13.103 & 13.479 & 0.0009 & 0.0010 & 0.0012 & 15.507 & 14.945 & 14.756 & 0.012 & 0.011 & 0.012 \\
    \textbf{45} & 15.494 & 15.455 & 15.619 & 0.0032 & 0.0032 & 0.0035 & 16.649 & 16.257 & 16.134 & 0.004 & 0.004 & 0.005 \\
    \textbf{46} & 14.287 & 14.358 & 14.590 & 0.0018 & 0.0019 & 0.0021 & 15.451 & 15.195 & 15.114 & 0.003 & 0.004 & 0.004 \\
    \textbf{47} & 14.384 & 14.461 & 14.696 & 0.0019 & 0.0020 & 0.0022 &       &       &       &       &       &  \\
    \textbf{48} & 14.117 & 14.256 & 14.488 & 0.0017 & 0.0018 & 0.0020 & 15.278 & 15.07 & 15.017 & 0.003 & 0.004 & 0.004 \\
    \textbf{49} & 13.839 & 13.966 & 14.211 & 0.0015 & 0.0016 & 0.0017 & 14.99 & 14.775 & 14.719 & 0.003 & 0.004 & 0.004 \\
    \textbf{51} & 14.512 & 14.469 & 14.646 & 0.0020 & 0.0020 & 0.0021 & 15.669 & 15.307 & 15.166 & 0.004 & 0.004 & 0.004 \\
    \textbf{53} & 14.025 & 13.328 & 14.267 & 0.0016 & 0.0012 & 0.0018 & 15.175 & 14.145 & 14.22 & 0.003 & 0.004 & 0.002 \\
    \bottomrule
    \end{tabular}%
  \label{table3}%
\end{table}%

\begin{figure}[!htb]
  \begin{center}
    \centering{\epsfig{file=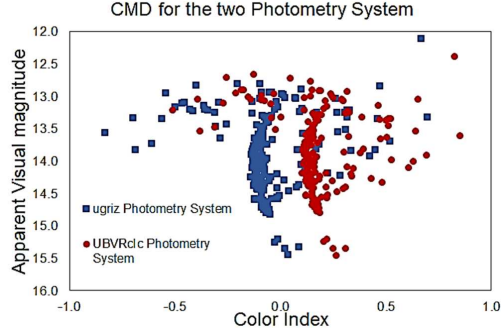, width=0.5\textwidth}}
    \caption[]{The color magnitude diagram for the $ugriz$ and $UBVR_cI_c$ standard photometry system}
    \label{fig5}
  \end{center}
\end{figure}

Further, these \textit{g}, \textit{r}, and \textit{i} magnitudes are the transformed into the \textit{B}, \textit{V}, and \textit{R} band magnitudes of Johnson-Cousins UBVRI standard photometric system by using transform equations: Eqs. (\ref{eq_17}) to (\ref{eq_19}) \cite{r22}. The calculated \textit{B}, \textit{V}, and \textit{R}, magnitudes are presented in the Table \ref{table4}. These calculated \textit{B}, \textit{V}, and \textit{R} magnitudes are compared with the NOMAD-1 catalog magnitudes \cite{r23}. Also, the color magnitude diagram (CMD) prepared from the calculated calibrated magnitudes of the 192 selected stars for both the standard photometry systems are presented in Fig. \,\ref{fig5}.

\begin{eqnarray}\label{eq_17}
     B &=& g_{cal} + 0.3130\times{(g_{cal} - r_{cal})} + 0.2271
  \end{eqnarray}
\begin{eqnarray}\label{eq_18}
     V &=& g_{cal} - 0.5784\times{(g_{cal} - r_{cal})} - 0.0038
  \end{eqnarray}
\begin{eqnarray}\label{eq_19}
     R &=& r_{cal} - 0.2936\times{(r_{cal} - i_{cal})} - 0.1439
\end{eqnarray}

\begin{table}[htbp]
  \centering
  \caption{Transformed \textit{B}, \textit{V}, and \textit{R} magnitudes}
    \begin{tabular}{cccccccccccccc}
    \toprule
    \multirow {2}{2.1em}{\centering{\textbf{Stars}}} &      \multicolumn {3}{p{6em}} {\centering{\textbf{Calculated Calibrated Magnitudes}}} &        
    \multicolumn {3}{p{6em}} {\centering{\textbf{Catalog Magnitudes}}} & 
    \multirow {2}{2.1em}{\centering{\textbf{Stars}}} &
    \multicolumn {3}{p{6em}} {\centering{\textbf{Calculated Calibrated Magnitudes}}} &        
    \multicolumn {3}{p{6em}} {\centering{\textbf{Catalog Magnitude}}}  \\
    \cmidrule(lr){2-7} \cmidrule(lr){9-14}   & 
      \multicolumn{1}{p{2em}}{\centering{\textit{\textbf{B}}}} & \multicolumn{1}{p{2em}}{\centering{\textit{\textbf{V}}}} & \multicolumn{1}{p{2em}}{\centering{\textit{\textbf{R}}}} & \multicolumn{1}{p{2em}}{\centering{\textit{\textbf{B}}}} & \multicolumn{1}{p{2em}}{\centering{\textit{\textbf{V}}}} & \multicolumn{1}{p{2em}}{\centering{\textit{\textbf{R}}}} &       & \multicolumn{1}{p{2em}}{\centering{\textit{\textbf{B}}}} & \multicolumn{1}{p{2em}}{\centering{\textit{\textbf{V}}}} & \multicolumn{1}{p{2em}}{\centering{\textit{\textbf{R}}}} & \multicolumn{1}{p{2em}}{\centering{\textit{\textbf{B}}}} & \multicolumn{1}{p{2em}}{\centering{\textit{\textbf{V}}}} & \multicolumn{1}{p{2em}}{\centering{\textit{\textbf{R}}}} \\      
    \midrule
    \textbf{8} & 13.882 & 13.766 & 13.752 & 14.44 & 14.54 & 14.39 & \textbf{31} & 13.855 & 13.656 & 13.590 & 14.23 & 14.50 & 13.99 \\
    \textbf{10} & 14.631 & 14.412 & 14.335 & 15.86 & 17.97 & 14.90 & \textbf{33} & 13.667 & 13.510 & 13.475 & 14.23 & 14.41 & 14.10 \\
    \textbf{13} & 13.720 & 13.332 & 13.372 & 14.30 & 14.22 & 13.78 & \textbf{34} & 12.976 & 13.063 & 13.186 & 13.46 & 17.97 & 11.77 \\
    \textbf{14} & 14.424 & 14.104 & 13.966 & 14.71 & 15.29 & 14.28 & \textbf{36} & 14.021 & 13.901 & 13.889 & 15.19 & 14.78 & 14.75 \\
    \textbf{16} & 13.334 & 13.219 & 13.329 & 14.09 & 14.08 & 13.99 & \textbf{37} & 13.873 & 13.357 & 13.290 & 14.52 & 14.21 & 13.90 \\
    \textbf{17} & 12.547 & 12.671 & 12.911 & 12.02 & 11.38 & 10.95 & \textbf{38} & 14.267 & 14.101 & 14.057 & 14.54 & 14.85 & 14.71 \\
    \textbf{20} & 13.563 & 13.387 & 13.527 & 16.37 & 17.78 & 13.06 & \textbf{40} & 14.743 & 14.584 & 14.543 & 15.02 & 14.62 & 15.33 \\
    \textbf{21} & 14.412 & 14.193 & 14.118 & 15.14 & 15.02 & 14.90 & \textbf{41} & 12.949 & 12.977 & 13.070 & 15.22 & 15.05 & 15.01 \\
    \textbf{22} & 14.721 & 14.112 & 13.800 & 15.13 & 14.89 & 13.83 & \textbf{45} & 15.734 & 15.468 & 15.359 & 16.27 & 15.99 & 16.35 \\
    \textbf{25} & 14.270 & 13.887 & 13.709 & 14.78 & 14.62 & 14.40 & \textbf{46} & 14.492 & 14.324 & 14.282 & 15.22 & 15.05 & 15.01 \\
    \textbf{26} & 14.727 & 14.415 & 14.281 & 15.31 & 15.21 & 15.10 & \textbf{47} & 14.587 & 14.425 & 14.386 & 15.23 & 15.22 & 15.07 \\
    \textbf{27} & 13.661 & 13.510 & 13.476 & 14.23 & 14.30 & 14.21 & \textbf{48} & 14.300 & 14.194 & 14.180 & 14.65 & 14.71 & 15.06 \\
    \textbf{28} & 14.526 & 14.392 & 14.369 & 15.17 & 15.27 & 15.18 & \textbf{49} & 14.026 & 13.909 & 13.894 & 14.44 & 14.86 & 14.22 \\
    \textbf{29} & 13.840 & 13.367 & 13.333 & 14.39 & 14.17 & 13.55 & \textbf{51} & 14.752 & 14.483 & 14.377 & 15.29 & 14.96 & 15.01 \\
    \textbf{30} & 13.354 & 13.060 & 13.169 & 13.99 & 13.94 & 13.73 & \textbf{53} & 14.470 & 13.618 & 13.460 & 15.10 & 14.42 & 13.93 \\
    \bottomrule
    \end{tabular}%
  \label{table4}%
\end{table}%

It is observed that there is certain deviation between the calculated and standard magnitudes. This is due to the accuracy with which the software counts the star minus background sky counts, accuracy of detecting the aperture center and the aperture parameters set while performing the aperture photometric analysis through Makali’i software. In addition, it is noted that the Makali’i softwar is capable of providing best results for the bright stars only. It is analyzed that the Makali’i software is effective for the stars which have star minus background sky count value of 500 or greater. For the validation of this statement, some of the dim stars are considered whose celestial coordinates are presented in Table \ref{table5}. In this analysis, the star celestial coordinates and their respective numbers are adopted from \cite{r19, r26}. The reason for selecting these stars is that most of these are the dim stars in the \textit{g}-band and are good candidates for dim star analysis by Makali'i software. On analyzing the \textit{g}-band images, it is observed that some dim stars have the result counts below 500 and for some stars the result counts are even zero. For these star the instrumental magnitude would not be accurate. Hence, it could be said that although the Makali'i software is simple and user friendly software but it is unable to detect and provide results counts for very faint stars. This software provides effective output for the bright stars.

\begin{table}[htbp]
  \centering
  \caption{Celestial coordinates for the analysis of dim stars in the \textit{g}-band images of NGC 2420}
    \begin{tabular}{cccccc}
    \toprule
    \multicolumn{3}{p{11em}}{\centering{\textbf{Undetectable Stars}}} & \multicolumn{3}{p{14.5em}}{\centering{\textbf{Stars having Result Count less than 500}}} \\
    \cmidrule(l){1-3} \cmidrule(l){4-6}    
    \multicolumn{1}{p{3em}}{\centering{\textbf{Star No.}}} & \multicolumn{1}{p{4em}}{\centering{\textbf{RA(\degree)}}} & \multicolumn{1}{p{4em}}{\centering{\textbf{DEC(\degree)}}} & \multicolumn{1}{p{2.5em}}{\centering{\textbf{Star No.}}} & \multicolumn{1}{p{6em}}{\centering{\textbf{RA(\degree)}}} & \multicolumn{1}{p{6em}}{\centering{\textbf{DEC(\degree)}}} \\
    \midrule
    \textbf{II} & 114.651 & 21.486 & \textbf{XV} & 114.834 & 21.335 \\
    \textbf{III} & 114.654 & 21.529 & \textbf{XVI} & 114.811 & 21.427 \\
    \textbf{V} & 114.492 & 21.529 & \textbf{XVII} & 114.821 & 21.546 \\
    \textbf{VI} & 114.501 & 21.452 & \textbf{XX} & 114.694 & 21.496 \\
    \textbf{VII} & 114.353 & 21.243 & \textbf{XXIII} & 114.382 & 21.230 \\
    \textbf{IX} & 114.393 & 21.279 & \textbf{XXV} & 114.214 & 21.695 \\
    \textbf{X} & 114.361 & 21.547 & \textbf{XXVI} & 114.374 & 21.682 \\
    \textbf{XIV} & 114.792 & 21.636 & \textbf{XXVII} & 114.681 & 21.967 \\
    \textbf{XXI} & 114.502 & 21.350 & \textbf{XXVIII} & 114.674 & 21.947 \\
    \textbf{XXII} & 114.535 & 21.444 & \textbf{XXIX} & 114.751 & 21.612 \\
    \textbf{XXIV} & 114.331 & 21.473 &       &       &  \\
    \bottomrule
    \end{tabular}%
  \label{table5}%
\end{table}%

\section*{Conclusion}
This article provides the insight into the physics of differential and all-sky aperture photometry by manually scanning the astronomical images. The astronomical images of NGC 2420 are analyzed through a Windows based software: Makali'i. The \textit{g}, \textit{r}, and \textit{i} magnitudes and error magnitudes of 192 selected stars present in an old open cluster NGC 2420 is calculated and then compared with the respective standard magnitudes and error magnitudes of the catalog. It is analytically observed that the Makali'i software does effective photometry analysis for the bright stars. However, for the faint star the software is unable to obtain the photon counts. It is noted that for count value (star minus background sky) of less than 500, the instrumental magnitude is not accurate. To overcome this deviation instead of using Makali'i software one could utilize aperture photometry tool (APT). Further, all the calibration parameters required for obtaining the calibrated magnitude is described in detail which are then used to obtain the calibrated instrumental magnitude. In addition, the calibrated instrumental magnitudes are transformed into the \textit{B}, \textit{V}, and \textit{R} band magnitudes of Johnson-Cousins UBVRI standard photometric system. Also, the color magnitude diagrams for both $ugriz$ and $UBVR_cI_c$ standard photometric systems are presented which are obtained from the calculated calibrated magnitude. Thus, this article aids the non-professionals to familiarize with the fundamentals of the differential and the all-sky aperture photometry. 

{\textbf{Acknowledgments:}} The authors are grateful to Dr. Rucha Desai (Associate Professor, PDPIAS, CHARUSAT), Dr. Dipanjan Dey (Assistant Professor, ICC, CHARUSAT) and Mr. Parth Bambhaniya (Research Scholar, ICC, CHARUSAT) for their valuable suggestions in the completion of this work. 


\end{document}